# DIRECT, a low-cost system for high-speed, low-noise imaging of fluorescent biological samples

ISABELL WHITELEY,[1,2,*] CHENCHEN SONG,[3] GLENN A HOWE,[1] THOMAS KNÖPFEL,[2,3] AND CHRISTOPHER J ROWLANDS[1,2]

[1]*Department of Bioengineering, Imperial College London, London, UK*
[2]*Centre for Neurotechnology, Imperial College London, London, UK*
[3]*Department of Brain Sciences, Imperial College London, London, UK*
*\*i.whiteley18@imperial.ac.uk*

**Abstract:** A targeted imaging system has been developed for applications requiring recording from stationary samples at high spatiotemporal resolutions. It works by illuminating regions of interest in rapid sequence, and recording the signal from the whole field of view onto a single photodetector, and can be implemented at low cost on an existing microscope without compromising existing functionality. The system is characterized in terms of speed, spatial resolution, and tissue penetration depth, before being used to record individual action potentials from ASAP-3 expressing neurons in an *ex vivo* mouse brain slice preparation.

## 1. Introduction

There are many advantages to using optical microscopy to investigate biological systems such as cells, tissues and superficial organs. It is fast, can resolve single cells and sub-surface features, and causes little damage to the sample under investigation. Nevertheless, cutting-edge biological applications have ever-more-stringent requirements in terms of speed, limits on photobleaching and tolerance of tissue scattering, all while maintaining the spatial resolution and field of view (FOV) that researchers have come to expect from their microscopes.

Achieving these performance improvements in the general case (for which the structure of the sample is unknown) is an enormous engineering challenge, but in a subset of applications, what matters is not the structure itself, but how it changes over time. This is particularly true in the case of fluorescent sensors, in which the cell does not move on the timescale of the experiment, but *does* change in emission wavelength, intensity or fluorescence lifetime. For example, neuroscientists increasingly use optical techniques to monitor calcium ion concentrations and even membrane voltage potentials in populations of neurons, by genetically-expressing a fluorescent probe and monitoring the resulting signal using a microscope. Many biological fields use Förster resonance energy transfer (FRET) sensors which work by non-radiative transfer of energy between a donor and an emitter fluorophore (or quencher) causing a change in fluorescence level. FRET sensors can respond very rapidly to external conditions [1]; they are widely used to monitor protein interactions and conformational changes [2], metal ions [3], metabolites and biomarkers [4] and for fluorescence lifetime imaging [5] in live *in vivo* samples. As many of these applications require long-term monitoring, they require optical techniques that can minimize photobleaching, improve temporal resolution (as high as ~1kHz in the case of some voltage sensors [6]) and reduce the impact of tissue scattering.

So high is the demand for high-speed, targeted systems that a number of techniques have recently been developed that are able to optically target features such as individual neurons within a labelled population, while also achieving improved spatial and/or temporal resolutions. Multiphoton excitation is commonly used in these applications; to increase laser efficiency and maximize the number of neurons targeted, some methods split a two-photon laser beam into beamlets using a spatial light modulator (SLM) [7,8] or with a microlens array [9] and then scan the beamlets across the sample using galvanometric (galvo) scanning mirrors and record the activity with a camera (EM-CCD [7] or sCMOS [8,9]). These scanning setups report reduced photodamage to the samples but are limited in their scanning rates due to the mechanical galvo mirrors and limited camera frame-rates. Other methods, such as two-photon FACED microscopy, also split the laser into beamlets (this time distributed into a tilted line), recording the fluorescence signal onto a photomultiplier tube sampling at hundreds of megasamples per second [10,11]. Nevertheless, the frame rate is still limited by the need to scan the tilted line of foci using a galvo mirror, necessitating a trade-off between temporal and spatial resolutions.

SLMs can also be used in combination with computer generated holography (CGH) directly (i.e. without galvo scanning). It has primarily been used in two photon optogenetic targeting [12,13], however these examples are limited in the number of targets that can be addressed in the same experiment, and the temporal resolution of the systems do not achieve the speed necessary for imaging fluorescent voltage sensors and other high-speed probes. SLM-based CGH was also used in single photon imaging of neurons containing a voltage dye [14]. Here the SLM was able to achieve high quality lateral (single cell) and axial (~10µm) confinement and reduced baseline fluorescence and photodamage. It recorded onto a sCMOS camera at high speed but required post-processing to remove potential temporal distortions caused by the high-speed imaging rates of the camera. Additionally, only a single target was recorded from during each experiment, and no solution was offered to scale to multiple targets. SLMs, through numerous methods, are able to generate spatially precise targeting, however they are temporally limited and costly to implement. Temporal focusing on the other hand can be used to precisely target thousands of individual neurons at sampling rates up to 160Hz [15]. Though it enables spatially-precise lateral and axial resolution, its limited temporal resolution of the highlighted system prevents its use in optical targeting experiments where there are rapid fluctuations in fluorescence in a densely populated or scattering sample, and furthermore, as the number of targets is increased, the temporal resolution must be further decreased. Though high-speed multiphoton targeting has been applied over large volumes, the lasers required are fragile and expensive, thus the use of single-photon excitation would be advantageous.

Spatially precise targeting has also been achieved in neurons with a digital micromirror device (DMD) for both optogenetic activation [16] and imaging with genetically encoded voltage indicators (GEVIs) [17]. In both situations, activation and imaging of neurons, the DMD provided precise spatial resolution and reduced photobleaching, but its high speeds were not utilized to increase temporal resolution. In both cases, a single frame was projected by the DMD to project to all targets, rather than switching between targets at high speed. Using a camera to record the activity of the targeted regions is the spatially and temporally limiting factor in GEVI imaging experiments. Only the neurons that the camera can see can be targeted and pixel binning is required to achieve high speed frame rates, that are fast enough to record high speed fluorescent changes.

Analyzing the techniques above there is a clear need for a low-cost method for recording changes in fluorescent probes with high sensitivity and ~500-1000Hz bandwidth, while ideally preserving the favorable photobleaching characteristics and tolerance to tissue scattering that existing methods provide, compared to simple wide-field microscopy. To address these limitations we present *DMD Imaging with Rapid Excitation and Confined Targeting* (DIRECT), a method for the high-speed monitoring of temporally-varying fluorescence signals based on strong structural priors. It uses a high-speed DMD to project patterns onto the sample in rapid sequence, recording the fluorescence emission from each projected pattern using a single point detector. The high speed of the DMD permits more than a dozen regions of interest (ROIs) to be probed with effective frame rates of more than a thousand measurements per second. This preserves the spatial resolution (ability to distinguish one cell precisely from its neighbor) while allowing high-bandwidth recording of fluorescence changes. DIRECT exhibits minimal photobleaching, significantly higher speeds, reduced read noise, has less onerous requirements for high-speed data storage, and increased tolerance to tissue scattering relative to conventional wide-field imaging or DMD targeting using a camera, while remaining low-cost, easy to implement and compatible with almost all existing microscope systems.

In this paper we characterize the performance of DIRECT, in particular the recording speed, resistance to sample scattering, and photobleaching reduction. We discuss the design decisions made in its creation, paying particular attention to the choice of detector. Finally, we demonstrate DIRECT on a number of test samples, culminating in the imaging of ASAP3-expressing neurons in an *ex vivo* mouse brain slice.

## 2. Results and Discussion

DIRECT has a number of advantages over conventional camera-based widefield imaging. Here we characterize the speed of recording, tolerance to photobleaching, and tolerance to scattering. We also present an analysis of the achievable resolution, as well as a comparison of different single-point detectors to establish which is most suitable for use in DIRECT. Finally we show how DIRECT has been used to record membrane voltage fluctuations in *ex vivo* mouse brain samples.

### 2.1 Design

DIRECT (Fig. 1) was designed to be integrated into standard upright microscopes using widely available, off-the-shelf parts and with minimal disruption to the existing optical pathways in the microscope. It fits into the infinity path of the microscope, between the microscope objective and the tube lens. A custom-machined interface part was used to fix DIRECT's optical pathway to the microscope. This part incorporates Olympus microscope dovetails with a Thor Labs removable filter cube holder (computer aided design (CAD) model available) but can be easily adapted to dovetails from other microscope manufacturers.

A DMD (a type of SLM consisting of an array of micromirrors) is placed conjugate to the sample plane in DIRECT's optical pathway. Binary masks of the desired targets are loaded and displayed by the DMD in quick succession (up to 22kHz frame rate) and projected onto the sample. DMDs provide advantages over other light shaping devices such as liquid crystal SLMs due to their fast refresh rates, (allowing for many target regions to be imaged sequentially) and relatively low cost. After the shaped light reflects off the sample, the resulting fluorescence information is captured by a single-point detector, such as a photodiode, photomultiplier tube (PMT), or silicon photomultiplier (SiPM); a camera is also available for more conventional imaging. A theoretical comparison of different detectors is given in the next section.

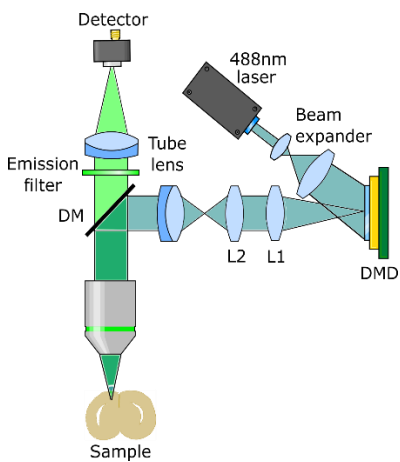

Fig. 1 Imaging pathway for targeted illumination. Light from a 488nm laser passes through a beam expander before striking a DMD. The reflected light is projected through a 4f lens system (L1 and L2) and enters the microscope through a tube lens. The light reflects off a dichroic mirror (DM) and through the objective onto the sample. The emitted light returns through the objective and DM, then passes through an emission filter and tube lens. The light is recorded by a detector (PMT, photodiode, SiPM, or camera). A removable cube containing a prism mirror is located after the beam expander to switch permit illumination with an LED if desired.

*2.2 Detector comparison*

DIRECT can use a variety of different detector types, including photodiodes, photomultipliers, silicon photomultipliers (SiPMs, also known as Multi-Pixel Photon Counters, or MPPCs), avalanche photodiodes and so on. Several of these were available in the laboratory, and their specified dynamic range, sensitivity and speed compared in order to establish their utility under different imaging conditions.

Neurobiological experiments in particular have exacting requirements; a very small change in fluorescence must be recorded at high speed (~1% change in the signal intensity at 1ms per frame [18]). Because fluorescence measurements are ultimately shot-noise limited, the probability distribution of the detected signal is given by a Poisson distribution, the signal-to-noise ratio (SNR) of which can be approximated by $\sqrt{m}$ (where $m$ is the number of detected events, detected photons in this case). Thus the detector must be able to capture at least 10,000 photons to achieve a SNR of 1:1 for a change in fluorescence ($\Delta F/F$) of 1%. Obviously, greater values are preferable.

To compare detectors under various conditions, it is necessary to know the expected photon count per pixel $F$ required to overcome the experimental noise. The total noise $n_{total}$ (defined as the mean deviation of a measurement from its true value) can be approximated as the combination of the detector dark noise and the photon shot noise of the overall measurement, summed in quadrature:

$$n_{total} = \sqrt{n_{shot}^2 + n_{dark}^2}$$

Eq. 1

Since the signal is expressed as a fractional change in fluorescence $\Delta F_\%$, the total number of fluorescence photons $F = n_{total}/\Delta F_\%$, for a SNR of 1. If the user desires a higher SNR, this expression must be multiplied by the relevant factor. Thus

$$F = SNR \frac{n_{total}}{\Delta F_\%}$$

Eq. 2

Finally, because $n_{shot}$ is the standard deviation of the shot noise of the measurement (which can be approximated as $n_{shot} = \sqrt{F}$), the following expression can be obtained:

$$F = SNR \frac{\sqrt{F + n_{dark}^2}}{\Delta F_\%}$$

Eq. 3

Which can be rearranged into the form of a quadratic:

$$\left(\frac{\Delta F_\%}{SNR}\right)^2 F^2 - F - n_{dark}^2 = 0$$

Eq. 4

This can be solved using the well-known quadratic formula

$$x = \frac{-b \pm \sqrt{b^2 - 4ac}}{2a}$$

Eq. 5

Substituting $x = F$, $a = (\Delta F_\%/SNR)^2$, $b = -1$ and $c = -n_{dark}^2$, and ignoring the negative solution resulting from subtracting the square root term:

$$F = \frac{1 + \sqrt{1 + 4 \times \left(\frac{\Delta F_\%}{SNR}\right)^2 \times n_{dark}^2}}{2 \times \left(\frac{\Delta F_\%}{SNR}\right)^2}$$

Eq. 6

For comparison, the same measurement can also be performed using a camera. The noise levels of a camera are slightly different though; because camera pixels have limited well depth $W$, and per-pixel read noise $n_{pix}$ is approximately constant regardless of integration time, many pixels must be summed together in order to reach the necessary photon counts; the total noise is therefore the sum of the noise from each pixel in quadrature. The number of pixels needed to capture a total of $F$ photons is, of course, heavily dependent on the spatial distribution of the fluorescence signal, but the lower bound is simply $F/W$. Consequently, a lower bound on $n_{dark}$ can be given as:

$$n_{dark} = \sqrt{\frac{F}{W} n_{pix}^2}$$

Eq. 7

For a camera, the quadratic expression therefore simplifies to

$$\left(\frac{\Delta F_\%}{SNR}\right)^2 F^2 - F - \frac{F}{W} n_{pix}^2 = \left(\frac{\Delta F_\%}{SNR}\right)^2 F^2 - \left(1 + \frac{n_{pix}^2}{W}\right) F = 0$$

Eq. 8

Ignoring the trivial solution $F = 0$, the required number of captured photons for a camera $F_{cam}$ is thus given by:

$$F_{cam} = \left(1 + \frac{n_{pix}^2}{W}\right)\left(\frac{SNR}{\Delta F_\%}\right)^2$$

Eq. 9

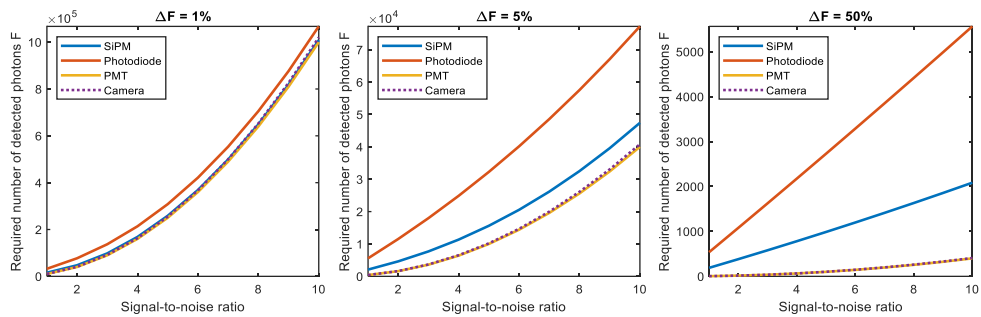

Fig. 2 Performance comparison between detectors under different imaging conditions. Simulated frame rate is 1000fps, and there are 10 independent illuminated areas. For cases where the fractional change in fluorescence is high, photon shot noise dominates and all detectors perform similarly. In cases where the fractional change is large however, detector read noise is significant and detector performance varies. Note also that the number of photons required to overcome photon shot noise decreases as the fractional change in fluorescence increases, as expected.

The results are summarized in Fig. 2; all cases describe a system in which integrated photon counts from 10 different areas must be recorded at 1000fps. A detailed description of each detector can be found in Table 1.

For cases where the fractional change in fluorescence $\Delta F_\%$ is low (on the order of 1%, consistent with early fluorescent voltage sensors like ArcLight [19]), there are no appreciable differences between detectors. When the fractional change in fluorescence reaches ~5%, the performance of the PMT and SiPM are broadly equivalent however there is a more pronounced performance deficit for the photodiode. Once the fractional change in fluorescence reaches 50% however (consistent with genetically-encoded calcium indicators such as jGCaMP7 [20]), the PMT has a clear performance advantage. Note that using a camera does not appear to improve sensitivity over that of a PMT (especially since real fluorescence distributions will be substantially more varied than the ideal case modelled), and in fact performs slightly worse than the PMT (although this performance deficit is reduced for higher $\Delta F_\%$).

**Table 1. Properties of selected detectors**

| Detector | Read noise | Maximum signal |
|---|---|---|
| **PMT: Hamamatsu H9305-03** | 3720 photons / s [a] | $6.24 \times 10^8$ photons / s, $10^5$ gain [b] |
| **Photodiode: Femto LCA-S-400K-SI-FST** | 2,680,000 photons / s | $4.29 \times 10^{12}$ photons / s [c] |
| **SiPM: Hamamatsu C13366-3050GA** | 937,000 photons / s | $12.6 \times 10^9$ photons / s |
| **Camera: Photometrics Kinetix** | 2 photons / pixel / frame | 200 photons / pixel / frame |

[a]Read noise calculated as dark current/ radiant sensitivity. [b]Maximum signal calculated as (max current/ charge of an electron)/ electron multiplication gain [c]Maximum signal calculated as max current/ charge of an electron.

### *2.3 Projection speed*

The number of ROIs that can be measured and the rate at which these measurements can be taken is controlled directly be the frame-rate of the DMD. The frame exposure time for each mask on the DMD was manually adjusted within the software used to run the experiments. The minimum exposure time at which the DMD was used was 100μs per frame (10kHz frame rate) as the DMD firmware became unstable at higher rates, though it is possible to achieve a frame rate of up to 22kHz. The inter-frame time (i.e. the time between the masks being displayed while the DMD was switching) was measured. At frame exposure times of 100μs, 1000μs, 10000μs, and 100000μs the inter-frame time between masks was 60μs. This time is included in the frame exposure times of the DMD. Over the course of a 10 second trial targeting 10 regions, and at a frame rate of 10kHz, each region is being targeted at a 1kHz rate and only 6% of the experimental time per sample is taken by the DMD switching between masks. This implies that fluorescence signals with a bandwidth of 500Hz (after considering Nyquist sampling) can be recorded, which comfortably exceeds the bandwidth of most fluorescent indicators in biological samples [6].

*2.4 Resolution and scattering tolerance*

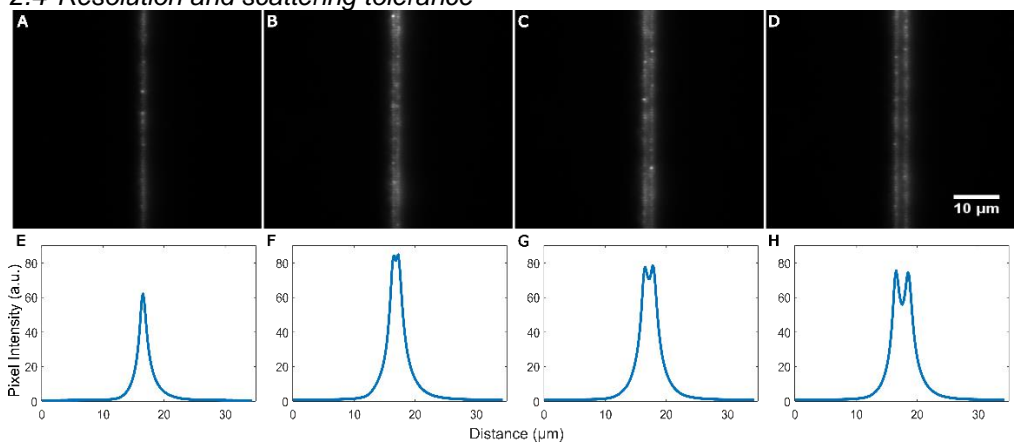

Fig. 3 Spatial resolution of DIRECT's targeting, A. a single pixel width line, B. a double pixel width line, C. two single pixel width lines with a line of off pixels in between, D. two single pixel width lines with two lines of off pixels in between. E-H. average intensity plots of the above lines; the mean of all the pixel values in each column are plotted to ascertain whether there is a reduction in intensity between the projected image of the mirrors.

Because the system was optimized such that the DMD image substantially covered the field of view (rather than to maximise spatial resolution at the cost of reduced field of view), the achievable resolution was limited by the size projected image of the DMD mirrors. This was first tested under ideal, non-scattering conditions. To test whether the relay optics were sufficient to maintain the resolution set by the DMD pixel size, a thin sample consisting of a monolayer of 100nm fluorescent beads was prepared. Single-mirror-width lines were projected onto the sample; images were captured of a single line, two adjacent lines, two lines with a row of off mirrors separating them, and finally two lines with two rows of off mirrors separating them (Fig. 3 A-D). The Sparrow limit was selected as the resolution criterion as it is more stringent than the more common Rayleigh criterion [21] while being simple and unambiguous to assess. The Sparrow limit is reached when the dip in light intensity between two points can no longer be resolved; it assumes uniform light intensity for both points. In the targeting resolution experiments, the two projected lines adjacent to each other, with no off mirrors in between are resolvable by the Sparrow criterion as the two separate intensity peaks are clearly visible (Fig. 3 B,F). The gap that is resolved is the gap between the mirrors of the DMD, thus we can conclude that the resolution of DIRECT is only limited by the mechanical design of the DMD.

The average pixel intensity of the projected lines was measured (Fig. 3 E-H), and the full-width, half-maximum (FWHM) of the plots was acquired and then compared to the theoretical FWHMs of a diffraction limited system. For a single row of mirrors, the FWHM was 1.69µm and the theoretical was 0.71µm at the sample plane. The acquired FWHMs for the other instances were 2.60µm, 3.23µm, and 3.80µm respectively. For the different images we found the FWHM of the acquired data to be on average 1.04µm±0.094µm wider than the theoretical FWHMs. We believe that much of the increase in width is due to light scatter from the fluorescent bead sample.

DIRECT was designed to be used for biological tissues and therefore needs to be able to target ROIs through densely scattering samples. Because all scattered photons are integrated onto a single detector (thus scattering of emission photons has negligible effect), DIRECT is expected to have heightened tolerance to tissue scattering compared to techniques like confocal microscopy which are sensitive to scattering of the emission photons. Nevertheless, because the optical properties of tissue samples are difficult to control, scattering phantoms, with controllable levels of light scattering, were used as a proxy. As in the above targeting resolution experiment, lines were projected by the DMD at different separation distances and imaged onto a thin layer of fluorescent microspheres. Above the fluorescent spheres, dilutions of Intralipid 20% were used as phantoms to represent different levels of scattering [22] (Fig. 4). The upright microscope's existing epifluorescence camera was supplemented by a second camera in a transmission geometry. Briefly, a custom-machined adaptor replaced the condenser, allowing an objective to be placed (in a geometry reminiscent of an inverted microscope) under the sample, imaging the fluorescent microspheres. An elliptical mirror in the adaptor enabled the light from the "inverted" objective to be projected through a tube lens onto a camera (Fig. 4). Images were taken from both epifluorescence and inverted cameras simultaneously, and their intensities were compared (Fig. 5). With low levels of scattering, projected lines in the transmission and reflection images were resolvable. As levels of scattering increased, the transmission images could still be resolved, however the reflection images were not resolvable. In densely scattering situations, DIRECT was still able to achieve precise targeting resolution even when the reflected image was not visible through the scattering. When used in combination

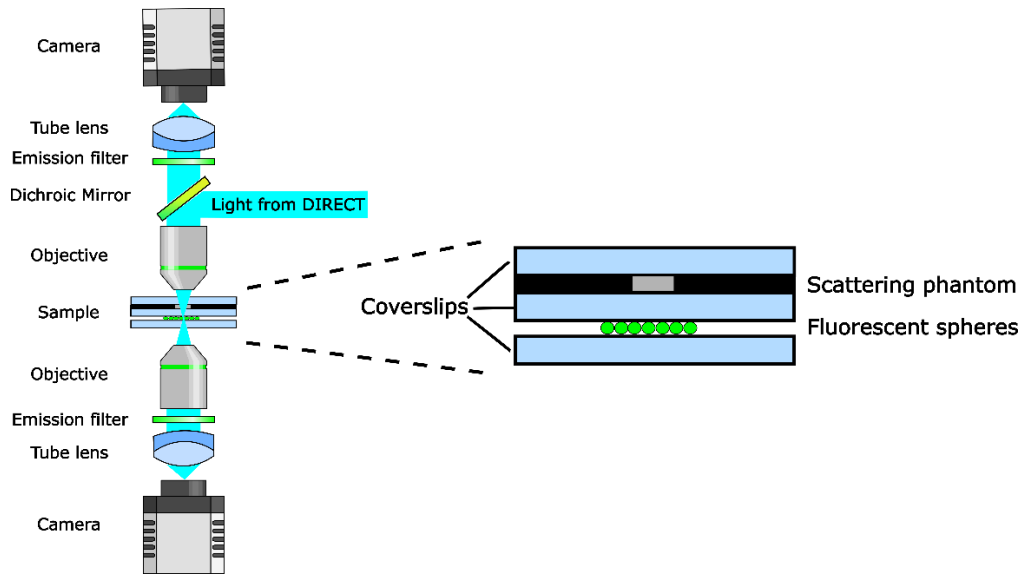

Fig. 4 Experimental design for scattering tolerance assessment. Left: upright and inverted microscopes. The inverted microscope is fitted onto the existing upright microscope at the location of the condenser lens using a custom milled objective holder with an Olympus condenser dovetail. Right: scattering sample design. From bottom up, coverslip, thin layer of fluorescent microspheres, coverslip, 200µm thick spacer with 5mm diameter well containing Intralipid 20% phantom, and coverslip.

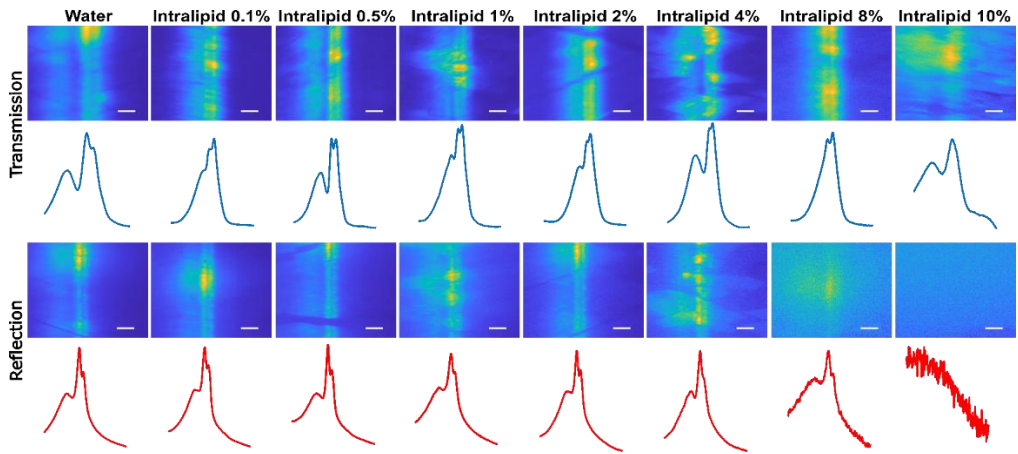

Fig. 5 Transmission and reflection images of two single mirror width lines, separated by two columns of off mirrors, projected through a scattering phantom. Dilutions of intralipid 20% were used to increase scattering density. Plots show the average pixel intensity of the projected lines. Scale bar = 5µm

with single-point detectors such as photodiodes or PMTs, DIRECT can accurately target and record from ROIs whose exact shape is unknown (see section 2.2).

*2.5 Photobleaching*

Photobleaching is a persistent issue in many biological experiments [23]. To reduce it, the light source power can be reduced, but this can lead to an increase in background noise. In a shot noise limited system, the noise scales as approximately the square root of the signal intensity, so decreasing the excitation power reduces the signal-to-noise ratio. Additionally, regions that are not actively being recorded from are still exposed to light and subject to photobleaching, which is damaging to the sample and can limit experimental utility. Because DIRECT does not illuminate cells that are not being recorded from, it should substantially reduce off target photobleaching.

Photobleaching rates for widefield fluorescence and DIRECT were compared. A series of images were captured over 20 minutes using both techniques: uniform illumination in the case of widefield fluorescence imaging, or switching between ROIs in the case of DIRECT. The average pixel value of each captured image or targeted region within the image was calculated and normalized to the exposure time, and then exponential decay curves were fitted to the averaged data (Fig. 6). At the same laser power and switching between three ROIs (~ 55 µm diameter circles) at 100 frames per second, DIRECT has an approximate twofold improvement in rate of decay compared to widefield imaging (decay rate of DIRECT = $-2.34\times10^{-3}$, decay rate of widefield = $-4.63\times10^{-3}$) (Fig. 6A,C). When the laser power was increased by a factor of three to account for the reduced amount of time DIRECT spends illuminating each target, we unexpectedly found that the decay rate was still approximately two times better than widefield (decay rate of DIRECT at increased power = $-1.98\times10^{-3}$)(Fig. 6B). Speculation on the mechanism of this improved tolerance to on-target photobleaching is beyond the scope of this paper, but nevertheless provides further support for the benefits of using DIRECT.

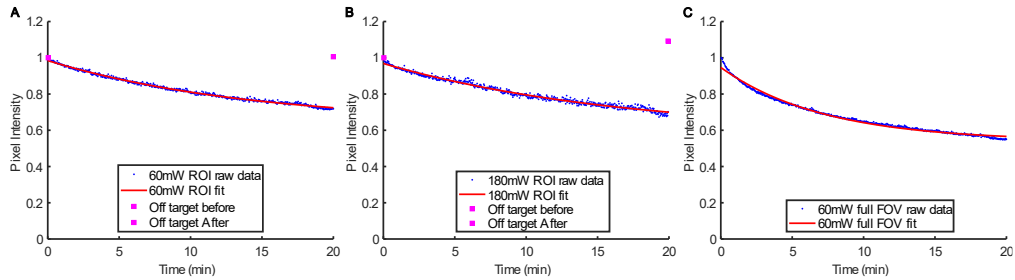

Fig. 6 Photobleaching decay rates using average pixel values of targeted regions with DIRECT versus full FOV widefield imaging. A. a three-ROI photobleaching decay curve at 60mW laser power, decay rate = $-2.342 \times 10^{-3}$, B. a three-ROI photobleaching decay curve at 180mW laser power, decay rate = $-1.976 \times 10^{-3}$, C. widefield photobleaching decay curve at 60mW laser power, decay rate = $-4.628 \times 10^{-3}$.

## 2.6 Voltage Imaging

While DIRECT can be used for a variety of applications, it is particularly suited to the high-speed high-dynamic-range measurements needed for imaging genetically-encoded fluorescent voltage indicators in brain tissue. For this demonstration, DIRECT was used to selectively target and record from multiple neurons simultaneously in an *ex vivo* mouse brain sample containing ASAP3-expressing neurons; ASAP3 is a genetically-encoded voltage sensor which changes fluorescence in response to changes in cell membrane potential [24]. While many genetically-encoded voltage sensors operate in this manner, typically the actual fluctuations are small and they are prone to bleaching [18]. Therefore, they require low background noise and high speed imaging to resolve the changes. The imaging system must also achieve a high spatial resolution to avoid bleaching the voltage indicator-containing neurons that are not being recorded from. DIRECT is an ideal system for recording from voltage indicators as its high speeds can capture the fluctuations in fluorescence and its targeting cuts background noise and off-target bleaching, while also reducing on-target photobleaching by decreasing the time the light spends on each target.

***DIRECT vs widefield imaging.*** Before recording data at high speed it is illustrative to assess just the effect of the targeted ROIs, using a camera rather than a single-point detector. ASAP3-expressing neurons across the field of view of the camera were targeted and a mask containing the selected neurons was generated. The resulting image of the projected mask was compared to a widefield image taken of the same FOV. The images were normalized such that the targeted neurons had the same average intensity and the background/non-targeted regions were compared (Fig. 7). In the WF image, the target intensity was negligibly higher than the background intensity (target to background ratio = 0.9293) whereas the ratio of intensity of the targeted region compared to the nontarget region of DIRECT was much larger (target to background ratio = 6.7440). While the intensity of the light inside the ROIs was similar, the intensity of the background of the widefield image was much greater than that of the ROI image. By using DIRECT, the background noise of the image was drastically reduced while the target intensity was unchanged. This allows for recordings of voltage activity with less noise while preserving other fluorescence-labelled neurons for further experimental recording.

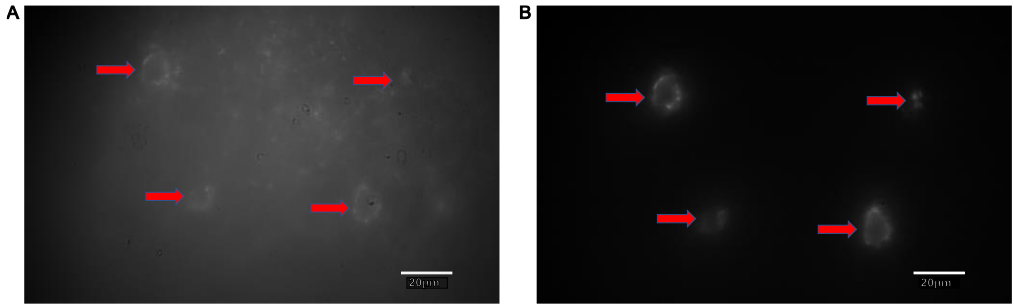

Fig. 7 Widefield vs DIRECT imaging. Representative example of widefield voltage imaging compared to DIRECTs targeted illumination. A. Widefield image of an in vitro sample with GEVI containing neurons, red arrows on widefield example indicate neurons selected for targeting. B. The same in vitro sample with targeted imaging of the selected neurons, demonstrating the removal of background fluorescence resulting from the use of targeted imaging.

*Fluorescence activity detection*. As a final demonstration, DIRECT was used to record electrically-induced action potentials in ASAP3-containing neurons. Masks of each targeted neuron were generated and projected onto the sample by the DMD for 100μs before switching to the next mask. The emitted photons were collected by a PMT. Ten neurons across the FOV were targeted (Fig. 8A) and the PMT recorded at 1MS/s; the high sampling rate enables accurate off-line synchronization of the measured data. There was an effective sampling rate of 1kHz per neuron from the DMD. An electrode was placed into the sample at a close proximity to the targeted neurons and a short electrical pulse was given, followed by 5 consecutive pulses 0.5 sec later. Activity traces of the 10 neurons were separated (Fig. 8B,E) and the ΔF/F and SNR ratio for the first action potential was calculated. The average SNR for the first action potential was 6.23 (σ = 1.59, n = 10 neurons) and the mean $\Delta F/F$ of the ASAP3 indicator during the first action potential was -7.62% (σ = 0.0098, n = 10 neurons) (Fig. 8C,D). The average intensity of each targeted neuron differed due to the different axial depths of each neuron, however the signal to noise ratio was high enough to resolve the changes in fluorescence that occurred due to electrical stimulation, even at the lowest baseline fluorescence values. By utilizing DIRECT, and switching between the different targets, the individual activity traces of each neuron could be easily separated and single action potentials for each target were resolved. DIRECT combined with a PMT enabled neurons at different depths to be recorded from simultaneously despite the precise shape of the neuron being unknown.

The total number of ROIs that can be recorded from during a single experiment without losing any activity information is determined by minimum sample rate that can be used, which in turn is determined by the rise and fall times of the fluorescence indicators used. In the above experiments, the system is able to unambiguously resolve signals with a bandwidth of 500Hz

(Nyquist sampling). If this sampling rate is retained, then DIRECT is able to target up to 22 ROIs (based on the DMD switching rate).

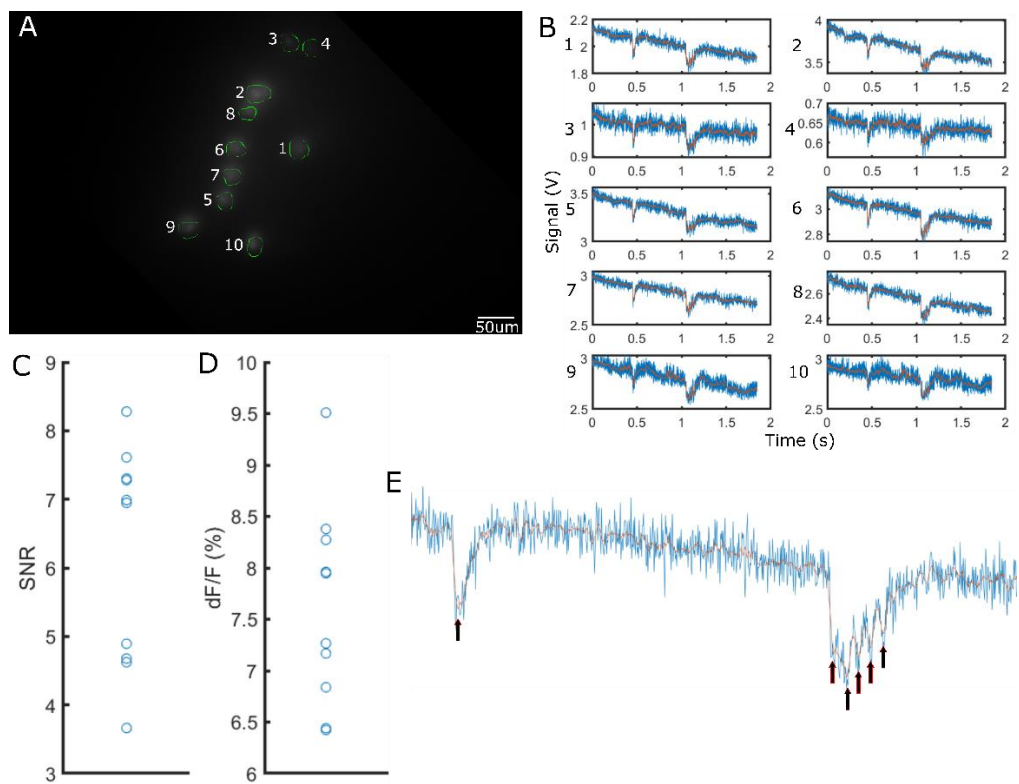

Fig. 8 Single trial multi-cell recording. A. Targeted illumination of the sample, green circles are the drawn ROIs for each neuron. B. Single trial traces of each ROI on left during extracellular stimulation experiment. Electrode was placed in proximity to the targeted neurons. A single pulse was given at approximately 0.5s into the experiment, followed by 5 pulses half a second later. Action potentials are visible at each pulse for each neuron. C,D. Signal to noise ratio and change in fluorescence for the first action potential of each neuron (n= 10). E. Close view example of single trial data in B, black arrows indicate voltage response to each extracellular electrical pulse. Blue is raw data, orange line is smoothed data using a moving average filter with a span of 10 data points.

## 3. Methods

### 3.1 Optical design

The DIRECT system is described as follows: light from a 488nm laser (Coherent Sapphire 488-200) passes through a 4.5× beam expander (Thorlabs A220TM-A and LBF254-050-A) and strikes a DMD (Vialux V-7001) located conjugate to the sample plane of the microscope. An alternative illumination path is also available, for light-emitting diode (LED) rather than laser illumination; after the beam expander but before the DMD, a mirror (Thorlabs MRA25-E02, placed in a removable cube: Thorlabs DFM1B-M and Thorlabs DFM1T2 cube insert) can be placed to switch to the LED.

The light reflected from the DMD passes through a 4f lens system (for characterization experiments: four Thorlabs LA1256-A lenses in Plössl pairs, for voltage imaging experiments: two Thorlabs LA1417-A lenses) onto an intermediate plane (where optionally a mask may be placed) and into a tube lens (Olympus SWTLU-C) before reflecting off a dichroic mirror (Semrock Di03-R488-t1-25x36 mounted in a Thorlabs DFM1/M removable filter cube, which is in turn mounted in a custom holder with Olympus dovetails) and through the objective (20x Olympus UPlanSApo for fluorescent bead experiments, either 16x Nikon CFI75 LWD 16X W or 60x Olympus LUMPLFLN60XW for voltage imaging experiments), onto the sample. The light emitted from the sample passes back through the dichroic and an emission filter (Semrock ff03-525/50-25), and is recorded by either a camera (Basler acA1920-155um or IDS U3-3080SE-M-GL), photodiode recording at 1kHz (Photonics 0210007), Silicon photomultiplier (SiPM, Hamamatsu C13366-3050GA) or PMT (Hamamatsu H9305-03). In all cases, signals from the non-camera detectors were digitized using an ADC module / oscilloscope (Pico Technology Picoscope 5444D).

DIRECT is intended to be inserted to the infinity path of a standard fluorescence microscope; in our demonstration we inserted it between the built-in fluorescence illuminator (Olympus BX-RFAA) and the trinocular head (Olympus U-TR30-2) of an Olympus BX51WI or BX61 using a custom-machined dichroic holder with Olympus dovetails. CAD models for all custom parts are available at https://www.imperial.ac.uk/rowlands-lab/; an optical diagram can be seen in Fig. 1. The system was designed and optimized using Autodesk Inventor Professional 2023 and Zemax OpticStudio.

### 3.2 Experimental control and ROI generation

The DMD allows arbitrary binary patterns to be projected onto the sample at high speed. This can be used to project individual ROIs. A custom software interface written in LabVIEW 2018 calling the functions from the Vialux ALP4.3 Dynamic Link Library (DLL) was used to upload binary sequences to the DMD; ROIs and patterns could be generated, and the rate of switching between each frame of the sequence could be selected for each experiment.

For an ROI projection experiment (where a sequence of ROIs was illuminated in quick succession), a widefield image was taken of the full FOV of the sample with no binning. The picture was loaded into the software interface and regions of interest were hand-drawn around the selected targets. Each region of interest was converted into a binary mask and uploaded to the DMD. To confirm accuracy of ROIs, all masks were combined into a single mask then projected onto the sample and a picture was captured.

### 3.3 Synchronization

Because the DMD could not be synchronized to the photodetector / ADC clock (or vice versa), synchronization between the two had to be performed offline. A start trigger was used to initialize the projection of the pattern sequence, after which the DMD and ADC board then ran asynchronously. Because the DMD was reset between frames (i.e. all mirrors returned to an 'off' position) the signal on the detector dropped to zero every frame, before rapidly increasing to a nonzero value, acting as a built-in clock. The ADC sampled the signal much faster than this clock, and could therefore observe the rapid increase and decrease. A custom MATLAB script was used to recover the clock (also compensating for slow drift between the DMD and ADC clocks); the script could optionally take an average of each 'on' signal for denoising and data reduction purposes.

### 3.4 Projection Speed

Masks of differing sizes were projected by the DMD at frame exposure times of 100μs, 1000μs, 10000μs, and 100000μs onto a fluorescent sample. Fluorescence was detected using a photodiode (Laser components Ltd Photoreceiver LCA-S-400K-SI) and oscilloscope (Picoscope 2204) sampling at 100kS/s. Mask switching was detected by identifying troughs between changing voltage levels as described previously.

### 3.5 Photobleaching

To assess on- and off-target photobleaching, a sample consisting of a close-packed array of 10× diluted 100nm fluorescent microspheres (Fluoresbrite YG Carboxylate Microspheres 0.10μm) was placed in the field of view. A reference image was taken of the whole field of view with all pixels illuminated. ROIs consisting of three ~55μm diameter circles were projected onto the sample in quick succession (10kHz), with the sequence repeating continuously throughout the experiment. Camera frames were taken continuously (exposure time 15ms, laser intensity at the sample 60mW over a ~425×350µm area) and the exposure continued for a period of 20 minutes. Finally a comparison image was taken of the whole field of view (once again with all pixels illuminated). The experiment was then repeated with the laser power increased by a factor of three (exposure time 15ms) (accounting for the threefold reduction in exposure duration caused by the sequential ROI exposure durations). A final experiment was done with all pixels illuminated for the whole duration of the experiment (exposure time 15ms) (simulating wide-field camera exposure). In each case a new, unexposed region of the sample was used.

### 3.6 Projection resolution and scattering tolerance

To assess the resolution of the targeting system under various imaging conditions, lines were projected by the DMD and imaged onto a camera (IDS U3-3080SE-M-GL) rather than a single point detector, in order to eliminate the effect of sample heterogeneity. Tests were performed by projecting two single-mirror-width lines onto a sample with rows of OFF mirrors between them. The sample was composed of a microscope slide with a thin layer of dried fluorescent microspheres (Fluoresbrite YG Carboxylate Microspheres 0.10µm) and a coverslip. The average pixel intensity was taken horizontally across the vertically projected lines to determine the intensity profile and separation of the lines.

The resolution of the targeting was assessed in controlled scattering conditions. An inverted microscope (Fig. 4) consisting of a microscope objective (Olympus UPlanApo 20x), elliptical mirror (Thorlabs BBE1-E02), tube lens (Thorlabs TTL200MP), emission filter (Chroma HQ535/50 x), and camera (IDS U3-3080SE-M-GL) was placed below the sample to acquire the transmission image of the pattern after passing through a scattering medium. The inverted microscope was fitted to the upright microscope setup at the location of the condenser. The condenser lens was removed from its mount and a custom milled dovetail-adaptor, holding the objective and elliptical mirror, was placed.

The above projection resolution experiment was repeated, this time through the scattering sample and both transmission and reflection images were captured. Scattering phantoms were created using dilutions of Intralipid 20% (Sigma-Aldrich 68890-65-3). Dilutions used were 0.1%, 0.5%, 1%, 2%, 4%, 8%, and 10%. The sample consisted of a thin layer of fluorescent microspheres (Fluoresbrite YG Carboxylate Microspheres 0.10µm) between two coverslips, a 200µm thick spacer with a 5mm diameter well containing a scattering phantom and a third coverslip on top (Fig. 4). Spacer was made using made 3M 9088 White Double Sided Plastic Tape; the well was made using a 5mm hole punch.

### 3.7 In Vitro slice preparation

Adult C57BL/6 mice were injected with 1µl of AAV9.mDlx-ASAP3-Kv into the somatosensory cortex. After 3 weeks expression time, mice were terminally anaesthetized with ketamine/xylazine, and transcardially perfused with ice-cold sucrose-based cutting solution (osmolarity of 300-310), with the composition (in mM): 3 KCl, 26 $NaHCO_3$, 1.25 $NaH_2PO_4$, 3 Na pyruvate, 0.5 $CaCl_2$, 4 $MgCl_2$, 190 sucrose, and 25 dextrose (pH7.4 bubbled with carbogen), and brain extracted. Coronal slices of 250µm thickness were cut with a Leica TS1200 vibratome and immediately transferred to holding artificial cerebrospinal fluid (ACSF, osmolarity of 300-310) at 34°C, with the composition (in mM): 126 NaCl, 3.5 KCl, 26 $NaHCO_3$, 1.25 $NaH_2PO_4$, 2 $CaCl_2$, 2 $MgSO_4$ and 10 dextrose (pH7.4 bubbled with carbogen), and allowed to recover for 30 min before transferring to room temperature. Recording ACSF (osmolarity of 300-310) was heated to 34°C and similar to holding ACSF in composition except with 1.2 $CaCl_2$ and 1 $MgSO_4$.

### 3.8 Neurophysiological activity recordings

***DIRECT vs Widefield*** To assess the functional use of DIRECT for biological applications in scattering tissue, the spatial resolution of DIRECT was compared widefield imaging in an *in vitro* mouse brain sample with neurons containing the voltage indicator ASAP3. Previous papers have also shown the reduction of background light using a DMD to target neurons [17], and this was also confirmed with DIRECT. A mask of four neurons was projected by the DMD and a widefield image of the same FOV was also captured onto a camera (Basler acA1920-155um).

***Neurophysiological Activity*** A widefield image was captured using the laser at 20mW and ROIs drawn around the targeted neurons as described in previous section. During the experiment, the laser power was set to 200mW.

Multi-ROI PMT recordings with extracellular stimulation: an extracellular electrode was placed in close proximity to the targeted neurons. Individual ROI masks of each target were loaded into the DMD, the experiment was triggered to begin. A single pulse was given by the electrode 0.5 seconds after the experiment began. It was followed by five consecutive pulses at one second intervals. The pulse current was 10-100µA with 200µs pulse duration. The five consecutive pulses were given at 20Hz. The DMD switched between all ROI masks in a continuous pattern with a mask exposure time of 100µs. Emitted photons were recorded by a PMT with an oscilloscope (PicoScope 5444D PC Oscilloscope 200 MHz 4 channel, Pico Technology) sampling at 1MHz for two seconds.

### 3.9 Data analysis

All system characterization, image, and neurophysiological data analysis was performed in Matlab R2022a.

***System Characterization***

*Projection speed.* Multi-ROI Picoscope data was converted to Matlab files, the troughs between ROIs were identified and the average time of the troughs was taken.

*Photobleaching assessment.* For ROI images, the mask used to project the ROIs was used to isolate the target regions of the sample, and the average value of targeted regions was taken. For full FOV images, the average value of each image gathered in the time series was taken for full FOV images. The average values of each image were normalized to the starting image value. Photobleaching was assumed to be monoexponential with an offset to account for camera pixel offsets. The time series means were therefore curve-fitted using a nonlinear least square method to the following equation:

$$f(x) = ae^{(-bx)} + c$$

Eq. 10

where a, b and c are fitted constants. For off target results, the non-targeted regions were isolated in the full FOV before and after images and the average was taken and normalized to the starting value. The change was calculated by subtracting the after from the before.

*Resolution and scattering tolerance.* Images were loaded, normalized, and lines were manually isolated. The average pixel intensity was calculated across rows of the image. FWHM was measured by calculating the half maximum and interpolating the width at the value. Theoretical mirror widths were calculated for a diffraction limited system. To model the imaging of the projected DMD patterns under diffraction-limited conditions, the imaging pipeline was simulated using a Fourier optics approach. A simulated PSF was constructed assuming a diffraction-limited optical system along with a grid of pixels based on the detector and DMD pixel widths. This pixel grid was sufficiently subsampled to facilitate the inclusion of intra-mirror gaps of the given DMD fill factor. Columns of pixels in this grid, representing columns of DMD mirrors, were subsequently 'illuminated' thereby generating the projected

DMD pattern. This projected pattern was then convolved with the PSF to simulate the final diffraction-limited image of the DMD pattern.

*Neurophysiological Activity*

Individual traces of neurophysiological data recorded onto the PMT were separated using a lab-built function. The function fits a square wave (with variable period and duty cycle) to the data to separate each ROI activity trace. The data from each segment separated by the square wave is averaged to a single data point and the points are placed into a vector for each ROI. Data is presented in raw format, or with a moving average smoothing filter with spans of 10 data points for multi-neuron PMT recordings. Action potentials for $\Delta F/F$ measurements were calculated by taking the $\Delta F/F$ of each value in a subset of the data, known to contain the first action potential and selecting the maximum resulting value. The SNR was calculated by taking the maximum $\Delta F/F$ and dividing by the standard deviation of the baseline signal.

## 4. Conclusion

In summary, a new targeted imaging system, DIRECT, has been developed for applications in which optical measurements must be taken at impractically-high frame rates across a wide FOV, without compromising spatial resolution. The system is simple to implement and can be retrofitted with only minor modification to almost any microscope, with minimal disruption to existing functionality. DIRECT can project patterns as small as a single mirror width of the DMD, allowing for precise targeting resolution. The speed of the DMD allows for target switching as fast as 22kHz, well above the bandwidth of most fluorescent biological probes; this was demonstrated by recording from ASAP-3 expressing cells in an *ex vivo* slice preparation.

**Funding.** IW is grateful for an EPSRC-funded studentship from the Center for Doctoral Training in Neurotechnology at Imperial College London. CJR acknowledges funding from EPSRC (EP/S016538/1), BBSRC (BB/T011947/1), Wellcome Trust (212490/Z/18/Z), Cancer Research UK (29694 and EDDPMA-May22\100059), the Royal Society (RGS\R2\212305), the Chan Zuckerberg Initiative (2020-225443 and 2020-225707) and the Imperial College Excellence Fund for Frontier Research.

**Acknowledgments.** The authors are grateful to Dr Debora Machado Andrade Schubert for proofreading and giving feedback.

**Disclosures.** The authors declare no conflicts of interest.

**Data availability.** Data is available upon request. CAD models are available at https://www.imperial.ac.uk/rowlands-lab/

## 5. References

1. Y. Gong, C. Huang, J. Z. Li, B. F. Grewe, Y. Zhang, S. Eismann, and M. J. Schnitzer, "High-speed recording of neural spikes in awake mice and flies with a fluorescent voltage sensor," Science **350**, 1361–1366 (2015).
2. M. J. Lohse, M. Bünemann, C. Hoffmann, J.-P. Vilardaga, and V. O. Nikolaev, "Monitoring receptor signaling by intramolecular FRET," Curr. Opin. Pharmacol. **7**, 547–553 (2007).
3. A. M. Hessels and M. Merkx, "Genetically-encoded FRET-based sensors for monitoring Zn2+ in living cells," Metallomics **7**, 258–266 (2015).
4. Mohd. Mohsin, A. Ahmad, and M. Iqbal, "FRET-based genetically-encoded sensors for quantitative monitoring of metabolites," Biotechnol. Lett. **37**, 1919–1928 (2015).
5. R. N. Day and M. W. Davidson, "Fluorescent proteins for FRET microscopy: Monitoring protein interactions in living cells," BioEssays **34**, 341–350 (2012).
6. Y. Bando, C. Grimm, V. H. Cornejo, and R. Yuste, "Genetic voltage indicators," BMC Biol. **17**, 71 (2019).
7. Y. Shao, J. Q. D.d.s, X. Peng, H. Niu, W. Qin, H. Liu, and B. Z. Gao, "Addressable multiregional and multifocal multiphoton microscopy based on a spatial light modulator," J. Biomed. Opt. **17**, 030505 (2012).
8. A. M. Packer, L. E. Russell, H. W. P. Dalgleish, and M. Häusser, "Simultaneous all-optical manipulation and recording of neural circuit activity with cellular resolution in vivo," Nat. Methods **12**, 140–146 (2015).
9. T. Zhang, O. Hernandez, R. Chrapkiewicz, A. Shai, M. J. Wagner, Y. Zhang, C.-H. Wu, J. Z. Li, M. Inoue, Y. Gong, B. Ahanonu, H. Zeng, H. Bito, and M. J. Schnitzer, "Kilohertz two-photon brain imaging in awake mice," Nat. Methods **16**, 1119–1122 (2019).

10. J.-L. Wu, Y.-Q. Xu, J.-J. Xu, X.-M. Wei, A. C. Chan, A. H. Tang, A. K. Lau, B. M. Chung, H. Cheung Shum, E. Y. Lam, K. K. Wong, and K. K. Tsia, "Ultrafast laser-scanning time-stretch imaging at visible wavelengths," Light Sci. Appl. **6**, e16196–e16196 (2017).
11. J. Wu, Y. Liang, S. Chen, C.-L. Hsu, M. Chavarha, S. W. Evans, D. Shi, M. Z. Lin, K. K. Tsia, and N. Ji, "Kilohertz two-photon fluorescence microscopy imaging of neural activity in vivo," Nat. Methods **17**, 287–290 (2020).
12. O. A. Shemesh, D. Tanese, V. Zampini, C. Linghu, K. Piatkevich, E. Ronzitti, E. Papagiakoumou, E. S. Boyden, and V. Emiliani, "Temporally precise single-cell-resolution optogenetics," Nat. Neurosci. **20**, 1796–1806 (2017).
13. A. Forli, M. Pisoni, Y. Printz, O. Yizhar, and T. Fellin, "Optogenetic strategies for high-efficiency all-optical interrogation using blue-light-sensitive opsins," eLife **10**, e63359 (2021).
14. A. J. Foust, V. Zampini, D. Tanese, E. Papagiakoumou, and V. Emiliani, "Computer-generated holography enhances voltage dye fluorescence discrimination in adjacent neuronal structures," Neurophotonics **2**, 021007 (2015).
15. R. Prevedel, A. J. Verhoef, A. J. Pernía-Andrade, S. Weisenburger, B. S. Huang, T. Nöbauer, A. Fernández, J. E. Delcour, P. Golshani, A. Baltuska, and A. Vaziri, "Fast volumetric calcium imaging across multiple cortical layers using sculpted light," Nat. Methods **13**, 1021–1028 (2016).
16. Y. Adam, J. J. Kim, S. Lou, Y. Zhao, M. E. Xie, D. Brinks, H. Wu, M. A. Mostajo-Radji, S. Kheifets, V. Parot, S. Chettih, K. J. Williams, B. Gmeiner, S. L. Farhi, L. Madisen, E. K. Buchanan, I. Kinsella, D. Zhou, L. Paninski, C. D. Harvey, H. Zeng, P. Arlotta, R. E. Campbell, and A. E. Cohen, "Voltage imaging and optogenetics reveal behaviour-dependent changes in hippocampal dynamics," Nature **569**, 413–417 (2019).
17. S. Xiao, E. Lowet, H. J. Gritton, P. Fabris, Y. Wang, J. Sherman, R. Mount, H. Tseng, H.-Y. Man, J. Mertz, and X. Han, "Large-scale voltage imaging in the brain using targeted illumination," 2021.04.05.438451 (2021).
18. T. Knöpfel and C. Song, "Optical voltage imaging in neurons: moving from technology development to practical tool," Nat. Rev. Neurosci. **20**, 719–727 (2019).
19. L. Jin, Z. Han, J. Platisa, J. R. A. Wooltorton, L. B. Cohen, and V. A. Pieribone, "Single action potentials and subthreshold electrical events imaged in neurons with a novel fluorescent protein voltage probe," Neuron **75**, 779–785 (2012).
20. H. Dana, Y. Sun, B. Mohar, B. K. Hulse, A. M. Kerlin, J. P. Hasseman, G. Tsegaye, A. Tsang, A. Wong, R. Patel, J. J. Macklin, Y. Chen, A. Konnerth, V. Jayaraman, L. L. Looger, E. R. Schreiter, K. Svoboda, and D. S. Kim, "High-performance calcium sensors for imaging activity in neuronal populations and microcompartments," Nat. Methods **16**, 649–657 (2019).
21. J. S. Silfies, S. A. Schwartz, and M. W. Davidson, "The Diffraction Barrier in Optical Microscopy," https://www.microscopyu.com/techniques/super-resolution/the-diffraction-barrier-in-optical-microscopy.
22. P. Di Ninni, F. Martelli, and G. Zaccanti, "Effect of dependent scattering on the optical properties of Intralipid tissue phantoms," Biomed. Opt. Express **2**, 2265–2278 (2011).
23. A. Diaspro, G. Chirico, C. Usai, P. Ramoino, and J. Dobrucki, "Photobleaching," in *Handbook Of Biological Confocal Microscopy*, J. B. Pawley, ed. (Springer US, 2006), pp. 690–702.
24. V. Villette, M. Chavarha, I. K. Dimov, J. Bradley, L. Pradhan, B. Mathieu, S. W. Evans, S. Chamberland, D. Shi, R. Yang, B. B. Kim, A. Ayon, A. Jalil, F. St-Pierre, M. J. Schnitzer, G. Bi, K. Toth, J. Ding, S. Dieudonné, and M. Z. Lin, "Ultrafast Two-Photon Imaging of a High-Gain Voltage Indicator in Awake Behaving Mice," Cell **179**, 1590-1608.e23 (2019).